\documentclass[aps,prl,reprint,groupedaddress]{revtex4-2}

\usepackage{amsmath}
\usepackage{amssymb}
\usepackage{graphics}
\usepackage{epsfig}
\usepackage{color}
\usepackage{lineno}

\begin{document}

\title{Three-dimensional topological insulator feature of ternary chalcogenide Ge$_2$Bi$_2$Te$_5$}

\author{Shangjie Tian$^{1, 2, 3, \dag}$, Yuchong Zhang$^{4, 5, \dag}$, Chenhao Liang$^{4, 5, \dag}$, Yuqing Cao$^{2, 3, \dag}$, Wenxin Lv$^{2, 3}$, Xingyu Lv$^{2, 3}$, Zhijun Wang$^{4, 5, *}$, Tian Qian$^{4, 5, *}$, Hechang Lei$^{2, 3, *}$, and Shouguo Wang$^{1, *}$}
\affiliation{
$^{1}$Anhui Provincial Key Laboratory of Magnetic Functional Materials and Devices, School of Materials Science and Engineering, Anhui University, Hefei 230601, China\\
$^{2}$School of Physics and Beijing Key Laboratory of Optoelectronic Functional Materials $\&$ MicroNano Devices, Renmin University of China, Beijing 100872, China\\
$^{3}$Key Laboratory of Quantum State Construction and Manipulation (Ministry of Education), Renmin University of China, Beijing 100872, China\\
$^{4}$Beijing National Laboratory for Condensed Matter Physics and Institute of Physics, Chinese Academy of Sciences, Beijing 100190, China\\
$^{5}$University of Chinese Academy of Sciences, Beijing 100049, China
}

\date{\today}

\begin{abstract}

The exploration of novel topological insulators (TIs) beyond binary chalcogenides has been accelerated in pursuit of exotic quantum states and device applications. 
Here, the layered ternary chalcogenide Ge$_2$Bi$_2$Te$_5$ is identified as a three-dimensional TI. 
The bulk electronic structure of Ge$_2$Bi$_2$Te$_5$ features a hole-type Fermi surface at Fermi level $E_{\rm F}$, which dominates the transport properties.
Moreover, an unoccupied topological surface state with a Dirac point located at 290 meV above $E_{\rm F}$ has been observed. 
Theoretical calculations confirm a bulk bandgap and a nontrivial $\mathbb{Z}_2$ topological invariant (000;1). 
The present study demonstrates that the material family of layered tetradymite-like ternary compounds is an important platform to explore exotic topological phenomena.

\end{abstract}

\maketitle

%\section{Introduction}
Topological insulators (TIs) are distinguished from conventional insulators by a massless Dirac cone surface state in the bulk energy gap, the so-called topological surface state (TSS), which are protected by the time reversal symmetry and remain gapless under a nonmagnetic perturbation \cite{Kane2005,Fu2007a,Fu2007b,Hasa2010,Moor2010,Kane2011,Qi2011,Bans2016}.
The spin orientation of the TSS is locked with respect to crystal surface momentum, resulting in a helical spin texture \cite{Kane2005,Fu2007a,Fu2007b}. 
TIs manifest many exotic phenomena, such as gapless helical states with Dirac-like dispersion on the surface or edge, and quantum spin Hall effect (QSHE), providing a fertile ground to realize new electronic phenomena, such as a magnetic monopole arising from the topological magnetoelectric effect and Majorana fermions at the interface with a superconductor \cite{Hasa2010,Moor2010,Kane2011,Qi2011}.
%The unique properties of TSS provide fertile ground to phenomena such as the quantum spin Hall effect  and the quantum anomalous Hall effect (QAHE) when broken time-reversal symmetry, both of which can hold dissipationless helical or chiral edge state \cite{Hasa2010,Qi2011}.
Thus, TIs attract increasing attention due to the novel topological quantum states which have potential applications in quantum computation and spintronics \cite{Fu2007b,Hasa2010,Moor2010}.
After the experimental confirmation of the QSHE in two-dimensional (2D) TI of HgTe/CdTe quantum wells \cite{Bern2006}, research expands rapidly to three-dimensional (3D) systems.
A family of binary chalcogenide semiconductors, such as Bi$_2$Se$_3$, Bi$_2$Te$_3$, and Sb$_2$Te$_3$, have been verified to be 3D TIs with conducting TSSs \cite{Chen2009,Hsie2009,Zhang2009a,Zhang2009b,Xia2009,Kuro2010,Hatc2011,Kim2011}.

The layered tetradymite-like ternary compounds with the general formula $mAX\cdot nB_2X_3$ (where $A$ = Ge, Sn, Pb, Mn; $B$ = Sb, Bi; $X$ = Se, Te) comprise the two types of homologous series of layered ternary compounds, $A_mB_2X_{3+m}$ ($m> 1, n=1$) and $AB_{2n}X_{3n+1}$ ($m=1, n\geq 1$). These layered compounds have complex structures compared to the binary counterparts, which have attracted significant research interest as promising thermoelectric materials \cite{Alak2021,Shel2000,He2017,Zhu2017} and phase-change materials \cite{Ovsh1968,Sun2007,Alak2021}. 
Furthermore, many of $AB_{2n}X_{3n+1}$ ($m=1, n\geq 1$) layered compounds have been identified as 3D TIs with TSSs experimentally, such as GeSb$_2$Te$_4$ \cite{Nurm2020}, GeBi$_2$Te$_4$ \cite{Okam2012,Neup2012,Li2021}, SnSb$_2$Te$_4$ \cite{Nies2014}, SnBi$_2$Te$_4$ \cite{Li2021,Frag2021}, PbSb$_2$Te$_4$ \cite{Soum2012,Kuro2012}, PbBi$_{2n}$Te$_{3n+1}$ ($m=1,n=1 - 3$) \cite{Soum2012,Erem2012,Okud2013,Papa2016,Paci2018} and PbBi$_4$Te$_4$Se$_3$ \cite{Shve2019}. 
In addition, the magnetic members of this material family MnBi$_{2n}$Te$_{3n+1}$ represent the first intrinsic magnetic TIs \cite{Zhang2019,Li2019a,Otro2019,Gong2019,Hao2019,Chen2019,Li2019b}. 
Theoretical calculations suggest that $A_2B_2X_5$ ($m=2, n=1$) are also TIs or magnetic TIs \cite{Sa2011,Sa2012,Kim2012,Li2023}.
In contrast to  $AB_{2n}X_{3n+1}$ ($m=1, n\geq 1$),  the experimental studies on the topological properties of $A_mB_2X_{3+m}$ ($m> 1, n=1$) materials are still scarce.

Theoretical predictions indicate that Ge$_{2}$Bi$_{2}$Te$_{5}$ is TI with a single Dirac cone located at $\Gamma$ point of Brillouin zone \cite{Kim2012}.
In this work, we successfully grew high-quality single crystals of Ge$_2$Bi$_2$Te$_5$. Electrical transport measurements reveal it shows a metallic behavior with dominant hole-type carriers. 
Angle-resolved photoemission spectroscopy (ARPES) measurements on Ge$_2$Bi$_2$Te$_5$ have identified a hole-type Fermi surface (FS), a Dirac point located at 290 meV above the Fermi level $E_{\rm F}$. 
This nontrivial band structure is further confirmed by theoretical calculations.
These results demonstrate that Ge$_2$Bi$_2$Te$_5$ is a 3D strong TI with a topological invariant of $\mathbb{Z}_2$ = (000;1). 

%\section{Methods}

\begin{figure}
	\includegraphics[scale=0.18]{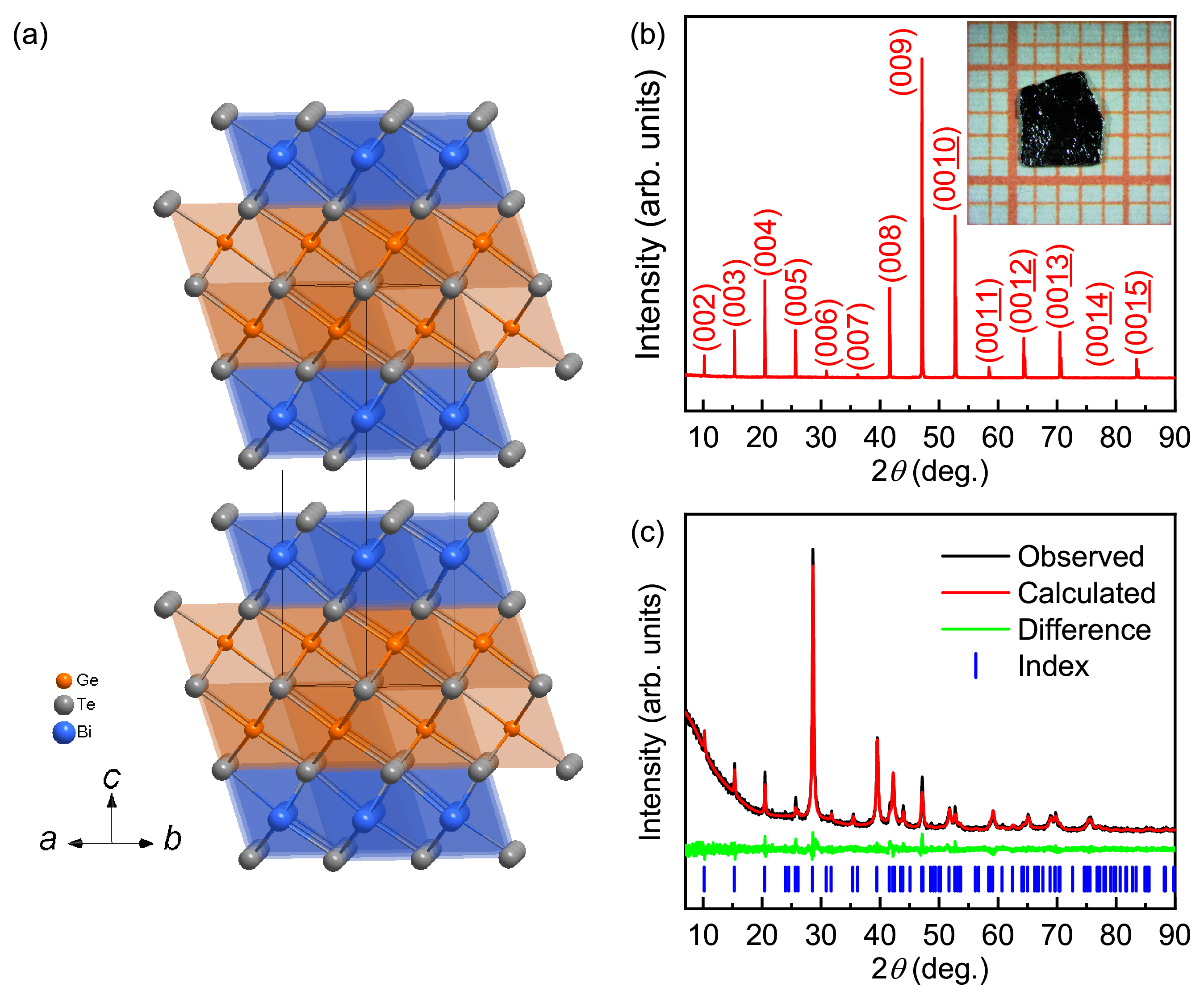}
	\caption{
		(a) Structure of Ge$_2$Bi$_2$Te$_5$ single crystal. The red, green and orange balls represent Ge, Bi and Te atoms, respectivety. 
		(b) XRD pattern of a Ge$_2$Bi$_2$Te$_5$ single crystal. The inset shows the photograph of a typical Ge$_2$Bi$_2$Te$_5$ single crystal.
		(c) Powder XRD pattern and Rietveld fit of crushed Ge$_2$Bi$_2$Te$_5$ crystals.}
	\label{FigStr}
\end{figure}

%\textbf{Crystal growth, structural and composition characterizations.}
Single crystals of Ge$_2$Bi$_2$Te$_5$ were grown by the self-flux method. The high-purity Ge (99.999 \%), Bi (99.999 \%) and Te (99.999 \%) shots were put into corundum crucibles and sealed into quartz tubes with a ratio of Ge : Bi : Te = 2 : 2 : 8. The tube was heated to $1273\ {\rm K}$ at a rate of $40\ {\rm K/h}$ and held there for $12\ {\rm h}$ to ensure a homogeneous melt. Then the temperature was rapidly cooled down to $672\ {\rm K}$ with subsequently cooling down to $593\ {\rm K}$ at $1\ {\rm K/h}$. The flux is removed by centrifugation, and shiny crystals with typical size about $2 \times 2 \times 0.1\ {\rm mm^3}$ can be obtained. The X-ray diffraction (XRD) patterns were performed using a Bruker D8 X-ray diffractometer with Cu $K_{\alpha}$ radiation ($\lambda = 0.15418\ {\rm nm}$) at room temperature. The elemental analysis was performed using energy-dispersive X-ray (EDX) spectroscopy analysis in a FEI Nano 450 scanning electron microscope.

%\textbf{Transport and magnetization characterizations.}
Electrical transport measurements were performed in a superconducting magnetic system (Cryomagnetics, C-Mag Vari-9). Both longitudinal and Hall electrical resistivity were measured simultaneously in a standard five-probe configuration with the current parallel to $ab$ plane and magnetic field along $c$ axis.
%The sample size used for transport measurement is $1.88 \times 1.96 \times 0.12\ {\rm mm^3}$ (length $\times$ width $\times$ thickness). 
In order to effectively get rid of the influence of voltage probe misalignment, we measured both resistivity at positive and negative fields. The final longitudinal and Hall resistivity were obtained by symmetrizing and antisymmetrizing raw data.

%\textbf{Angle resolved photoemission spectroscopy.}
ARPES experiments were conducted using a Scienta Omicron DA30-L analyzer.
The conventional ARPES was conducted at temperature 20K with probe pulse ($h\nu$ = 7.2 eV).
During the pump-probe measurement, the samples were pumped by an ultrafast laser pulse ($h\nu$ = 1.2 eV) with a pulse duration of about 200 fs and a repetition rate of 500 kHz. An ultraviolet probe laser pulse ($h\nu$ = 7.2 eV) subsequently photoemitted electrons. The overall time and energy resolutions were set to 300 fs and 10 meV, respectively. All pump-probe data are collected at a time delay of 0 ps. 
All the samples prepared for ARPES were cleaved in-situ and measured in an ultrahigh vacuum with a base pressure better than $1 \times 10^{-10}$ Torr. The polarization of probe pulse for both pump-probe and conventional measurement is $p$ polarization (containing both in-plane and out-of-plane component)

%\textbf{Theoretical Calculations.}
The first-principles band-structure calculations were carried out within the density functional formalism as implemented in VASP \cite{Kres1993, Kres1996} and use the all-electron projector augmented wave (PAW) \cite{Bloc1994, Kres1999} basis sets with the generalized gradient approximation (GGA) of Perdew, Burke, and Ernzerhof (PBE) \cite{Perd1996} for the exchange correlation potential. The Hamiltonian contains the scalar relativistic corrections, and the spin-orbit coupling was taken into account by the second variation method \cite{Koel1977}. Cutoff energy for the plane wave expansion was 620 eV and a $k$-point mesh of 12 $\times$ 12 $\times$ 3 is used for the bulk calculations. The experimental lattice parameters are employed for theoretical calculations.
	
%\section{Results and Discussion}

\begin{figure}
	\includegraphics[scale=0.17]{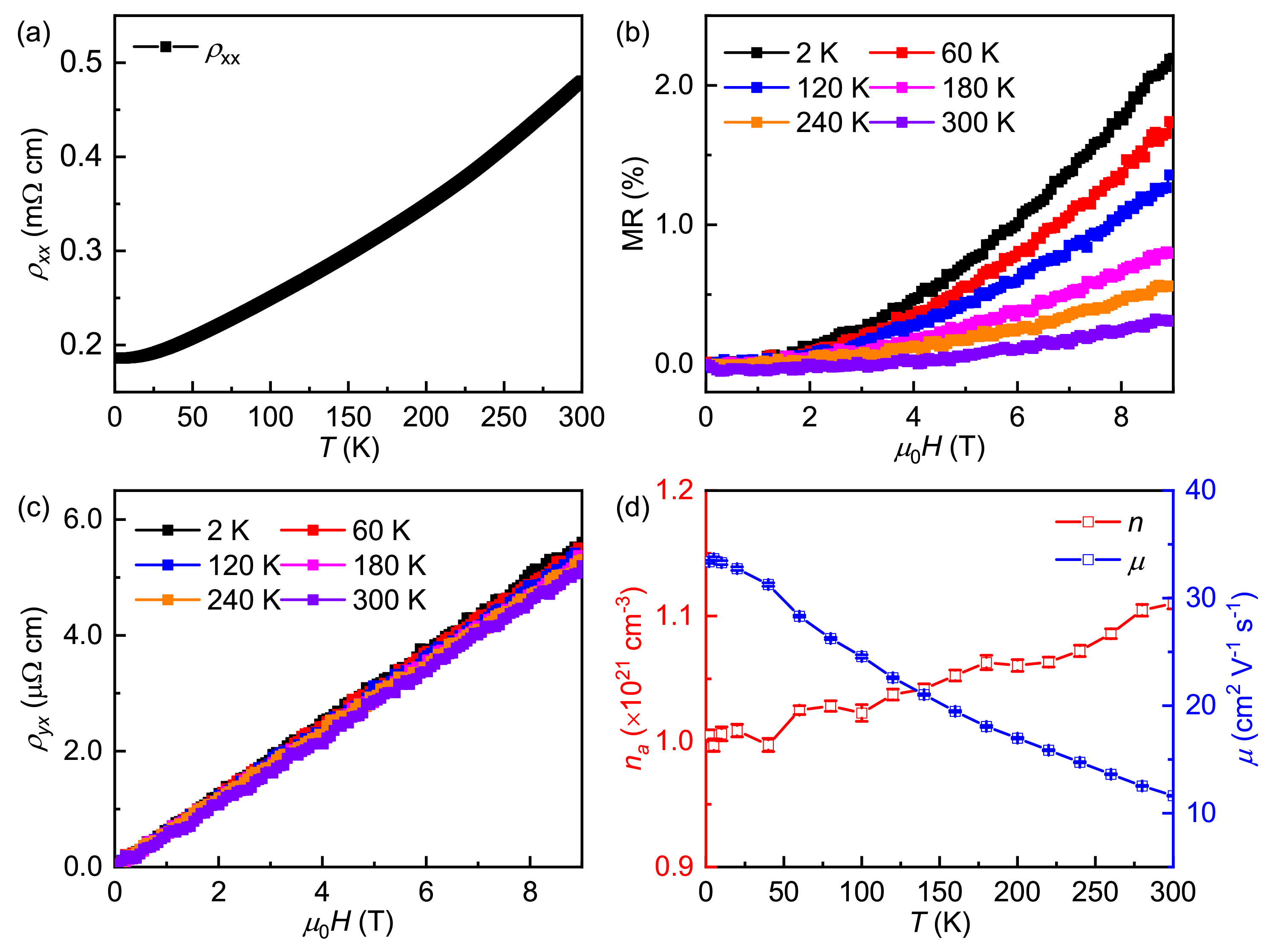}
	\caption{
		(a) Temperature dependence of $\rho_{xx}(T)$ of Ge$_2$Bi$_2$Te$_5$ single crystal. 
		(b) and (c) MR and Hall resistivity as a function of $\mu_{0}H$ between 2 K and 300 K. 
		(d) Temperature dependences of $n_a(T)$ and $\mu (T)$ derived from the linear fits of $\rho_{yx}(\mu_{0}H)$ curves and zero-field $\rho_{xx}(T)$ data.}
	\label{FigTran}
\end{figure}
	
\begin{figure*}
	\includegraphics[scale=0.21]{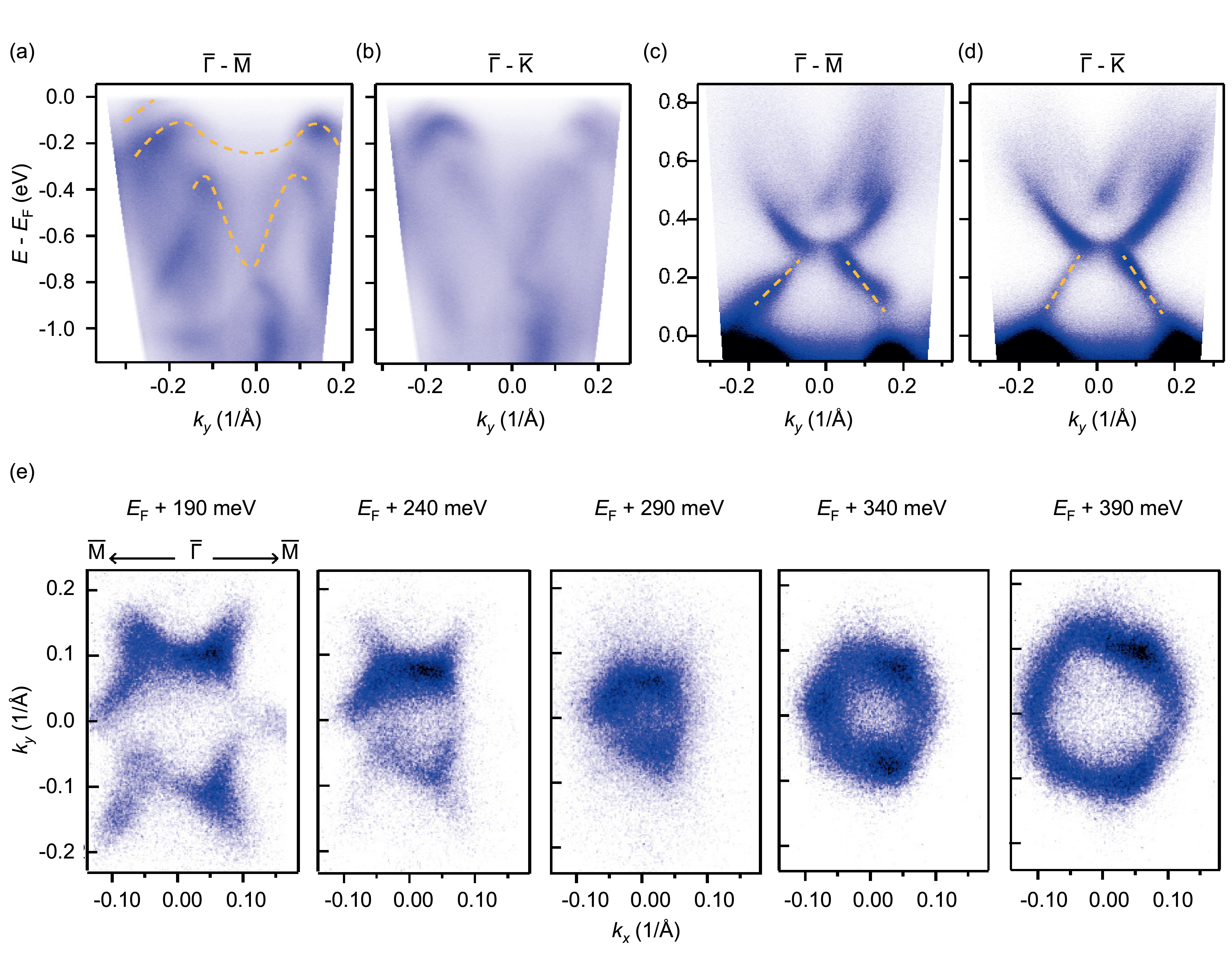}
	\caption{(a) and (b) Intensity plot of conventional ARPES data along $\bar{\Gamma}-\bar{\rm M}$ and $\bar{\Gamma}-\bar{\rm K}$ direction. Dashed lines (schematic) are guides to the eye indicating the dispersion of the bulk valence bands. (c) and (d) Intensity plot of pump-probe ARPES data along $\bar{\Gamma}-\bar{\rm M}$ and $\bar{\Gamma}-\bar{\rm K}$ direction. The dashed lines indicate the topological surface states. (e) Plots of the experimental constant energy contours (CECs) at different energy close to the Dirac point.}
	\label{FigARPES}
\end{figure*}

\begin{figure}
	\includegraphics[scale=0.25]{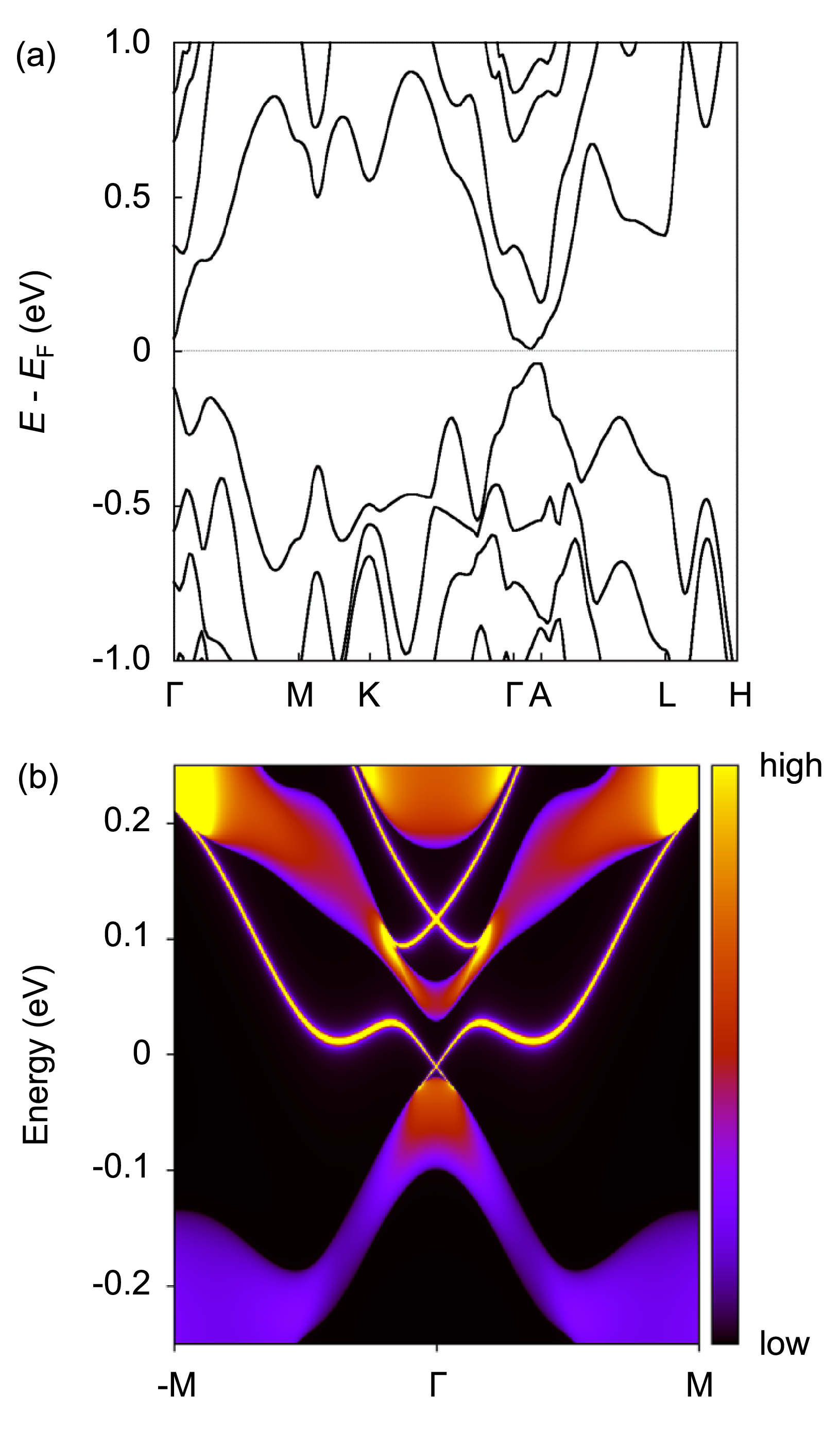}
	\caption{
		(a) Band structure of bulk Ge$_2$Bi$_2$Te$_5$ with spin-orbit coupling.
		(b) Surface state of the topmost Te--Bi--Te atomic trilayer at the Te-termination of Ge$_2$Bi$_2$Te$_5$.}
	\label{FigCal}
\end{figure}

As shown in Fig. \ref{FigStr}(a), Ge$_2$Bi$_2$Te$_5$ has a layered rhombohedra crystal structure (space group $P\bar{3}m1$, No. 164) \cite{Shel2000, Karp2000, Alak2021}. 
The basic building block of  Ge$_2$Bi$_2$Te$_5$  is the  nonuple layers (NLs). Within each NL, atoms are arranged in the stacking sequence Te-Bi-Te-Ge-Te-Ge-Te-Bi-Te. These NLs are stacked along the $c$-axis via weak van der Waals interactions.
Figure \ref{FigStr}(b) displays the XRD pattern of a Ge$_2$Bi$_2$Te$_5$ single crystal. All the peaks can be indexed by $(00l)$ reflections, which verifies the exposed surface of grown Ge$_2$Bi$_2$Te$_5$ being $ab$ plane and confirms the high quality of grown Ge$_2$Bi$_2$Te$_5$ crystals.
The photograph of a Ge$_2$Bi$_2$Te$_5$ crystal is presented in the inset of Fig. \ref{FigStr}(b). The plate-like morphology is consistent with the layered structure of Ge$_2$Bi$_2$Te$_5$. 
Figure \ref{FigStr}(c) displays the powder XRD pattern of crushed Ge$_2$Bi$_2$Te$_5$ crystal. It can be fitted very well using the reported rhombohedra structure in the literature \cite{Alak2021}. %The refinement parameters ($R_{wp} = 7.28 \%$) demonstrate high model reliability. 
The fitted lattice parameters are $a = b =$ 0.4286(2) nm and $c =$ 1.7368(1) nm,  in agreement with the results in the literature \cite{Shel2000, Karp2000, Alak2021}.
The actual atomic ratio determined from EDX measurement is Ge : Bi : Te = 1.91(2) : 2.11(2) : 5 when the Te content set as 5. It slightly deviates from the stoichiometric composition of Ge$_2$Bi$_2$Te$_5$, indicating the possible existence of antisite defects between Ge and Bi sites, which are commonly observed in this family of materials \cite{Shel2000}.

Figure \ref{FigTran}(a) shows the temperature dependence of longitudinal resistivity $\rho_{xx}$ of Ge$_2$Bi$_2$Te$_5$ crystals. 
The $\rho_{xx}$ decreases with the decrease of temperature $T$ from 300 K to 2 K, indicating the metallic behavior of Ge$_2$Bi$_2$Te$_5$. 
Figures \ref{FigTran}(b) and \ref{FigTran}(c) depict the dependence of the magnetoresistance (MR = $\frac{\rho_{xx}(\mu_{0}H) - \rho_{xx}(0)}{\rho_{xx}(0)}$) and Hall resistivity $\rho_{yx}(\mu_{0}H)$ of Ge$_2$Bi$_2$Te$_5$ at various temperatures. 
Across the entire measurement range (2 -- 300 K, 0 -- 9 T), Ge$_2$Bi$_2$Te$_5$ displays a weak positive MR ($<$ 3 \%). 
All the $\rho_{yx}(\mu_{0}H)$ curves exhibit positive slopes, indicating the dominance of hole-type carriers in the transport properties. 
By using the linear fits of the $\rho_{yx}(\mu_{0}H)$ curves and zero-field $\rho_{xx}(T)$ data, the temperature dependences of apparent carrier concentration $n_a(T)$ and mobility $\mu(T)$ can be determined based on single-band model ($R_{\rm H} = \rho_{yx}(\mu_{0}H)/\mu_{0}H$, $n_a= 1/eR_{\rm H}$,  and $\mu=R_{\rm H}/\rho_{xx}$, where $R_{\rm H}$ is Hall coefficient and $e$ is elementary charge). 
As shown in Fig. \ref{FigTran}(d), the $n_a$ of Ge$_2$Bi$_2$Te$_5$ at 2 K is $\sim 1.0 \times {10}^{21}\ {\rm {cm}^{-3}}$ and increases slightly with increasing temperature. 
Meanwhile, the $\mu$ decreases gradually from 33 cm$^2$ V$^{-1}$ s$^{-1}$ at 2 K to 12 cm$^2$ V$^{-1}$ s$^{-1}$ at 300 K. 
%The above results indicate that Ge$_2$Bi$_2$Te$_5$ is a typical hole-type metal.
This observation is consistent with previous reports that Ge$_{m}$Bi$_2$Te$_{3+m}$ ($m>1$ and $n=1$)  exhibit  hole-type behavior, while those GeBi$_{2n}$Te$_{3n+1}$ with $m=1$ and $n>1$ exhibit electron-type behavior. For GeBi$_2$Te$_4$ ($m=1$ and $n=1$), the dominant carriers are hole-type at low-temperature region and change to electron-type at high-temperature region \cite{Shel2000}. 
The dominant carriers are hole-type in GeTe, and when introducing Bi$_2$Te$_3$ layers, the formed antisite defects Bi$_\text{Ge}$ (Bi atoms occupy at Ge sites) contribute to the electron-type carriers \cite{Shel2000}. With increasing $n$, the carrier type evolves from hole to electron.

To further confirm the nontrivial topological property of Ge$_2$Bi$_2$Te$_5$, we turn to the measurement of electronic structure. The band structure along $\Gamma$--M and $\Gamma$--K is measured by conventional ARPES, as shown in Figs. \ref{FigARPES}(a-b). 
%\textcolor{red}{The bulk valence bands are consistent well with calculation (Fig. \ref{FigCal}(a)).}
The highest-lying band crosses $E_{\rm F}$ forming a hole-type FS and two M-shaped bands can be observed at higher binding energy, as indicated by the dashed curve. Since the sample is highly hole-doped and conventional ARPES measurement cannot access to the unoccupied states, we employed pump-probe setup to investigate the electronic states in the bulk gap above $E_{\rm F}$, where a Dirac-cone like TSS is expected to exist for $\mathbb{Z}_2$ TIs. ARPES intensity plots measured at zero time-delay presented in Figs. \ref{FigARPES}(c-d) show that a TSS crossing the bulk band gap appears near $\Gamma$ point, as indicated by the orange dashed line. The Dirac point locates at about 290 meV above $E_{\rm F}$. The band structure below $E_{\rm F}$ shows linear dispersion, and the Dirac velocity is estimated to be about 1.34 eV \AA$^{-1}$ for $\bar{\Gamma}-\bar{\rm M}$ direction and 1.83 eV \AA$^{-1}$ for $\bar{\Gamma}-\bar{\rm K}$ direction. 
The linear dispersion above Dirac point was not clearly observed in present experiment. It could be attributed to the fact that the linear dispersion above Dirac point exist in small energy range according to calculated band structure and that the energy resolution of trARPES measurements is limited. 
The distinction in Dirac velocity can also be observed in the constant energy contour plot. A bulk conduction band with higher binding energy can be seen in the energy range $E-E_{\text{F}}>0.4\ {\rm eV}$ near $\Gamma$ and the states with high spectral weight near $E_F$ indicate the bulk valence band. Fig. \ref{FigARPES}(e) shows the evolution of the Dirac-cone like dispersion of the TSS in the constant energy contours (CECs) at different binding energies. The CECs below Dirac point show hexagonal symmetry and the spectral weight distribution can be attribute to matrix element effect, while at energy of Dirac point, the CEC shows 3-fold symmetry. On the other hand, the CECs above Dirac point are circular, with three spectral weight peaks along the circle.

The theoretically calculated band structure is shown in Fig. \ref{FigCal}(a), where the calculated bulk gap is $\sim$ 50 meV. Considering the possible underestimation of GGA for the $s-p$ gap, we have improved the calculations by HSE method \cite{Kruk2006}, and still found the band inversion around 50 meV. We then use the IRVSP to analyze the irreducible representations of the electronic states at several high-symmetry $k$ points \cite{Gao2021}. Accordingly, the weak and strong indices are obtained by the parity criterion (000;1), indicating a strong TI phase. The topological Dirac states are expected on the (001) surface. For this purpose, we construct the tight-binding Hamiltonian with Ge-$p$, Bi-$p$, and Te-$p$ orbitals via the Wannier90 software package \cite{Most2008}. We use an iterative method to obtain the surface Green’s function of the semi-infinite system, as employed in the WannierTools \cite{Wu2018}. The imaginary part of the Green’s function is the local density of states (LDOS) at the surface. Exfoliation of Ge$_2$Bi$_2$Te$_5$ occurs at the Te-Te van der Waals interface. The surface states on this termination are shown in Fig. \ref{FigCal}(b).

%\section{Conclusion}

In summary, we establish the phase-change material Ge$_2$Bi$_2$Te$_5$ as a 3D strong TI through a comprehensive combination of transport measurements, ARPES experiments, and theoretical calculations.
ARPES measurements reveal a hole-type FS, consistent with the hole-dominated transport behavior observed in transport measurements.
The TSS and Dirac point at 290 meV above $E_\text{F}$ with anisotropic dispersion and 3-fold symmetric contours are observed via time-resolved ARPES.
First-principle calculations further confirm a band inversion exists in Ge$_2$Bi$_2$Te$_5$, yielding $\mathbb{Z}_2$ = (000;1) invariant. 
This nontrivial topological properties, combined with the established phase-change properties, highlight Ge$_2$Bi$_2$Te$_5$ as a unique multifunctional platform bridging topological quantum states and nonvolatile memory technologies.

\vspace{2mm}

%\begin{acknowledgments}
	
\textit{Acknowledgments}—This work was supported by the National Key R\&D Program of China (Grants Nos. 2022YFA1403800 and 2023YFA1406500), and the National Natural Science Foundation of China (Grant Nos. 12274459, 1218810, 12525409, 52130103 and U22A6005), and the China Postdoctoral Science Foundation (Grant No. 2023M730011), and the Center for Materials Genome.

%\end{acknowledgments}
	
$^{\dag}$ These authors contributed equally to this work.

$^{\ast}$ Corresponding authors: Z.J.W. (wzj@iphy.ac.cn); T.Q. (tqian@iphy.ac.cn); H.C.L. (hlei@ruc.edu.cn); S.G.W. (sgwang@ahu.edu.cn)

%\bibliography{Ge2Bi2Te5_Tran}

\end{document}